\documentclass[aps,showpacs,preprint]{revtex4-1}
\usepackage{graphicx}

\newcommand{\beq}{\begin{equation}}
\newcommand{\eeq}{\end{equation}}
\newcommand{\beqa}{\begin{eqnarray}}
\newcommand{\eeqa}{\end{eqnarray}}
\newcommand{\eps}{\epsilon}
\newcommand{\rr}{{\bf r}}

\newcommand{\s}{\sigma }
\newcommand{\p}{{\bf p}}
\newcommand{\Q}{{\bf Q}}
\newcommand{\V}{{\bf V}}
\begin{document}

\title{Spin-charge ordering induced by magnetic field in superconducting state: analytical solution in
the two-dimensional self-consistent model }
\author{S. I. Matveenko }
\affiliation{
L. D. Landau Institute for Theoretical Physics, 14243  Chernogolovka, Russia}
\author{S. I. Mukhin}
\affiliation{Theoretical Physics and Quantum Technologies Department, Moscow Institute for Steel and Alloys, 119049 Moscow, Russia. }

\date{\today}

\begin{abstract}
Solutions of the Bogoliubov-de Gennes equations
for the two-dimensional self-consistent Hubbard t-U-V model of superconductors with
$d_{x^2-y^2}$ symmetry of the order parameter in the presence of a magnetic field are found.
It is shown that spatial inhomogeneity of superconducting order parameter  results in the emergence of stripe-like
 domains that are stabilized  by  applied magnetic field leading to emergence of space-modulated composite
  spin-charge-superconducting order parameter.
\end{abstract}

\maketitle

\section{Introduction}
The accumulating experimental evidence suggests that the pseudogap phase could be a key
issue in understanding the underlying mechanism of high-transition-temperature
superconducting (high-T$_c$) copper oxides \cite{taill}. Magnetotransport data in electron-doped
copper oxide La$_{2-x}$Ce$_x$CuO$_4$ suggests that linear temperature-dependent
resistivity correlates with the electron pairing and spin-fluctuating scattering of the electrons \cite{greene}.
Simultaneously, in the hole-doped copper oxide YBa$_2$Cu$_3$O$_y$ a large in-plane
anisotropy of the Nernst effect sets in at the boundary of the pseudogap phase and indicates that this
phase breaks four-fold rotational symmetry in the a-b plane, pointing to stripe or nematic order
\cite{taill1}. Antiferromagnetic spin ordering in single crystals of underdoped ($x = 0.10$) La$_{2-x}$Sr$_x$CuO$_4$
 induced by an applied magnetic was observed in magnetic neutron diffraction experiments \cite{aeppli2}. The authors found that
 applied magnetic field, that imposed the Abrikosov's vortex lattice, also induced `striped' antiferromagnetic order with the ordered
moment  $0.4\mu_B$ per Cu$2^+$, as suggested by muon spin relaxation data \cite{muon}. The field-induced signal increased according
to classical mean-field theory, suggesting an intrinsic mechanism \cite{aeppli2}. Besides, The field-induced signal proved to be
resolution-limited, and the magnetic in-plane correlation length ($\zeta > 400 A$ ), was much greater than the superconducting coherence
length ($\xi<20 A$ ) and the inter-vortex spacing $a_v = 130A$  for H$ = 14.5$T. The authors concluded that since
 superconducting coherence length $\xi$ defines the size of the vortices, coherent magnetism scaled
  with $\zeta$ must extend beyond the vortex core across the superconducting regions of the material, and
  hence, superconductivity and antiferromagnetism coexist throughout the bulk of the material \cite{aeppli2}.
Motivated by these experimental results, we study here theoretically an analytically solvable model that manifests
 a coexistence of the stripe order and d-wave superconductivity.
We found an exact solution of the Bogoliubov-de Gennes equations in the simple microscopic Hubbard
t-U-V model indicating that the Abrikosov's vortex core naturally gives rise to a stripe-ordered domain, i.e. coupled
 antiferromagnetic (AFM) electronic spin- and charge-density modulations. We show that the size of the stripe domain may
  exceed superconducting vortex's core size $\xi_s$
and the inter-vortex distance in the Abrikosov's lattice in the limit of weak magnetic fields $H \leq H_{c1}$.
As far as we know, this is the first analytic solution of such type. Previously, coexistence of the
stripe-order and Abrikosov's vortices in the limit of high magnetic fields $H\sim H_{c2}$ has been investigated
numerically \cite{knapp}. Calculations were limited by the size of the model cluster of 26$\times$52 sites.
 Hence, the numerical results covered only the case when the inter-vortex distance was less than correlation
  length of the static AFM order. Here we consider analytically the opposite case of
weak magnetic fields, where the inter-vortex distance is mach greater than stripe-order (including AFM) correlation length.
\\

Predicted numerically stripe order \cite{Zaanen}, e.g. coupled spin- and charge-density periodic superstructure
(SDW-CDW), was found in the underdoped superconducting
cuprates experimentally, specifically in $La_{2-x}Ba_xCuO_4$ \cite{fujita} and
$La_{1.6-x}Nd_{0.4}Sr_xCuO_4$ \cite{ich,hu}.
It was shown analytically, that stripe-order may arise already in the short-
range repulsive Hubbard model in the form of  spin-driven CDW: there occurs an enhancement of the quantum
 interference between backward and Umklapp scattering of electrons on the SDW potential close to half-filling
  when a CDW order with "matching" wave-vector is present \cite{sm}.  In the quasi-1D case analytical
kink-like spin- and charge-density coupled solutions  were found \cite{mm} in the normal state. A study of
$La_{1.875}Ba_{0.125}CuO_4$ with angle-resolved photoemission and scanning tunneling
spectroscopies \cite{valla} has provided evidence for a d-wave-type gap at low temperature, well within the
stripe-ordered phase but above the bulk superconducting Tc. An earlier  inelastic neutron scattering
 data \cite{aeppli1} had shown field-induced fluctuating magnetic order with space periodicity $8a_0$ and
wave vector pointing along Cu-O bond direction in the ab-plane of the optimally doped
La$_{1.84}$Sr$_{0.16}$CuO$_4$ in the external magnetic field of $7.5$ T below $10$ K. The applied
magnetic field ($\sim 2-7$ T) imposes the vortex lattice and induces "checkerboard" local density of
electronic states (LDOS) seen in the STM experiments in high-T$_c$ superconductor
 Bi$_2$Sr$_2$Ca Cu$_2$O$_{8+\delta}$ \cite{davis}. The pattern originating in the Abrikosov's vortex cores
  has $4a_0$ periodicity, is oriented along  Cu-O bonds, and has decay length $\sim 30$ angstroms reaching
   well outside the vortex core.
The existence of antiferromagnetic spin fluctuations well outside the vortex
cores is also discovered by NMR \cite{nmr} in superconducting YBCO in a $13$ T external magnetic field.
Theoretical predictions had also been made of the magnetic field induced
coexistence of antiferromagnetic ordering phenomena and superconductivity in high-T$_c$
cuprates \cite{zhang,arovas,sachdev1,sachdev2,sachdev3} due to assumed proximity of pure superconducting state
 to a phase with co-existing superconductivity and spin density wave order. In these works effective Ginzburg-Landau
  theories of coupled superconducting-, spin- and charge-order fields were used.
Alternatively, the fermionic quasi-particle weak-coupling approaches were
focused on the theoretical predictions arising from the model of BCS
superconductor with $d_{x^2-y^2}$ symmetry \cite{volovik}. An effect of the
nodal fermions on the zero bias conductance peak in tunneling studies was
predicted. However STM experiments of the vortices in high-T$_c$ compounds
revealed a very different structure of LDOS \cite{davis}.
In this paper we make an effort to combine both theoretical approaches and
present analytical mean-field solutions of coexisting spin-, charge- and
superconducting orders derived form microscopic Hubbard model in the
weak-coupling approximation. The  previous analytical results obtained in
the quasi 1D cases \cite{mm,mm1,ma,fe} are now extended for two real
space dimensions. Different analytical solutions for collinear and checkerboard
stripe-phases, as well as for spin-charge density modulation inside
Abrikosov's vortex core are obtained. Simultaneously, our theory provides
wave-functions of the fermionic states in all considered cases.

\section{ The model.}

Consider   the Hamiltonian \cite{fe, fe2}

\beqa
H =\displaystyle -t\sum_{\langle i,j\rangle,
\sigma}c^{\dagger}_{i,\sigma}c_{j,\sigma}+ U\displaystyle\sum_{i}
\hat{n}_{i,\uparrow} \hat{n}_{i, \downarrow}
- \mu \sum_{i. \sigma} \hat{n}_{i, \sigma} + \nonumber\\
( \sum_{<i,j>, \s} \Delta(i,j;\sigma)c^{\dagger}_{i,\sigma}
c^{\dagger}_{j,-\sigma} + h.c.),
  \label{hubbard}
\eeqa
where
the first term is the kinetic energy,  the second  term is the  on-site repulsion
$U > 0$,
and the last ones  describe  a superconducting correlations.
A sum $\sum_{<i,j>, \s}$ is taken over nearest neighboring sites
${\bf r}_i$, ${\bf r}_j$ of the square lattice, and spin  components $\s  = \pm 1$.

In a self-consistent approximation\cite{mm}  we define
 slowly varying functions for spin order parameter
$m(\rr_i )$
and the charge density $\rho (\rr_i)$:
\beqa
\rho (\rr) = \langle \hat{n} (\rr ) \rangle, \quad
(-1)^{x_i + y_i} m (\rr_i ) =
- U \langle \hat{S}_z (\rr_i ) \rangle, \nonumber \\
\Delta(i,j;\sigma)= -g \langle c_{j,-\sigma}
c_{i,\sigma}\rangle.
\label{selfc}
\eeqa
Then  the Hamiltonian (\ref{hubbard})  takes a form
\beqa
H = -t \sum_{\langle i,j\rangle
\sigma}c^{\dagger}_{i,\sigma}c_{j,\sigma}  +\frac{U}{2}
\sum_{i,\sigma}(\rho_i c^{\dagger}_{i,\sigma}c_{i,\sigma} -\frac{\rho_i^2}{2})
\nonumber \\
-U \sum_{i, \sigma} \{\langle \hat{S}_z (\rr_i ) \rangle \sigma
 c^{\dagger}_{i,\sigma}c_{i,\sigma} -\langle \hat{S}_z (\rr_i ) \rangle^2\}
-\mu \sum_{i, \sigma}c^{\dagger}_{i,\sigma}c_{i,\sigma} \nonumber \\
\sum_{<i,j>,\s} \Delta(i,j;\sigma)c^{\dagger}_{i,\sigma}
c^{\dagger}_{j,-\sigma} +h.c. + \frac{|\Delta |^2}{g},
\label{H}
\eeqa

We  diagonalize this  Hamiltonian  using   Bogoliubov transformations
\beq
\hat{c}_{\sigma}(\rr) = \sum_n \gamma_{n, \sigma} u_{n, \sigma}(\rr)-\sigma
\gamma^+_{n, -\sigma} v^*_{n, -\sigma}(\rr),
\label{tr}
\eeq
with   new fermionic operators $\gamma$, $\gamma^+$.

We suppose the $d_{x^2 -y^2}$ symmetry of the
 superconducting order parameter $\Delta(\rr ,\rr \pm \hat{\bf x};\s)=
 \s \Delta_d (\rr )$, $\Delta(\rr ,\rr \pm \hat{\bf y};\s)=
 -\s \Delta_d (\rr )$ so that
 $\Delta_{\p} (\rr) = 2(\cos p_x - \cos p_y ) \Delta_d (\rr).
$
Consider states near the Fermi surface (FS)  (see Fig. 1)
 and use linear
approximation for the quasiparticles spectrum.
 We write the functions $u(\rr )$ and $v(\rr )$ as
\beq
u_{\s}(\rr ) =\sum_{\p \in FS, p_x >0}[u_{\p, \s} e^{i \p \rr}
+ \s u_{\p - \Q, \s} (\rr ) e^{i(\p -\Q )\rr }],
\eeq
where $\Q = \Q_+$ for wave vectors $\p_y > 0$ and $\Q = \Q_-$ for
$p_y < 0$, respectively.

 \begin{figure}[tbph]
\begin{center}
\includegraphics[width=3.0in]{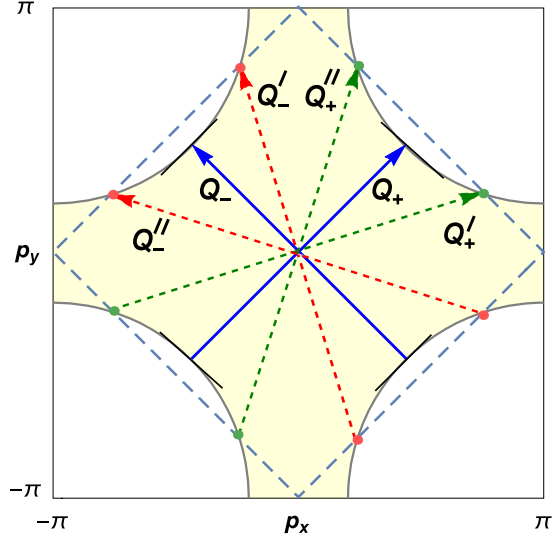}
\caption{(Color online) The Fermi surface in the Brillouin zone for a nearly half-filled squared lattice Hubbard model, with two independent nesting vectors $Q_{\pm}$ defined in the text. }
\end{center}
\label{fs}
\end{figure}

In the general case of a  doped system vectors $\Q_{\pm}$ differ from half-filled nesting vectors $\{\pi,\pm\pi\}$ in the reciprocal cell units, and become independent, so we rewrite
 $(-1)^{x_i + y_i} m (\rr_i ) $  as
\[
 m_+ (\rr_i )\exp (i \Q_+
\rr_+ ) + m_- (\rr_i )\exp (i \Q_- \rr_- ) + h.c.
\]
Note, that we consider for simplicity a two pairs  of almost plane sections of the Fermi surface with
 one pair of nesting vectors
 ${\bf Q}_{\pm}$  (See Fig.~1). In  real systems
there are two pairs of nesting vectors   ${\bf Q}^{\prime}_{\pm}$  (shown in Figure), and ${\bf Q}^{\prime \prime}_{\pm}$,
connecting opposite "hotspots".

In the continuous approximation  eigenvalue equations
 take the form: $\hat{H}\Psi = \eps \Psi$, with
\beq
  \hat{H} =
\left(
\matrix{-iV_{\p}\nabla_{\rr} +\eps_{\p} -\eta&
 m_{\pm} (\rr) & \Delta_{-\p} &0 \cr
m^*_{\pm} (\rr)  & -iV_{\p-\Q}\nabla_{\rr} +\eps_{\p -\Q} -\eta
 & 0& \Delta_{-(\p -\Q)}\cr
 \Delta_{\p}^*  &0 & iV_{\p}\nabla_{\rr} -\eps_{\p}+
\eta & m_{\pm} (\rr) \cr
 0 & \Delta_{\p -\Q}^* & m^*_{\pm}(\rr ) & iV_{\p
-\Q}\nabla_{\rr}
 -\eps_{\p -\Q}+ \eta }
\right),
\label{HH}
\eeq
where $\eta = \eta(\rr) = \mu - \frac{U}{2}\rho(\rr )$,
$\Psi^T = (u_{\p}, u_{\p-\Q}, v_{\p}, v_{\p-\Q}) =(u_+, u_-, v_+, v_- )$,
$\eps_{\p} = -2t(\cos p_x + \cos p_y ) -\mu$,
${\bf V}_{\p}=2 t (\sin p_x, \sin p_y )$.
The vector  $\Q = \Q_+$ for  $\p_y > 0$, and $\Q = \Q_-$ for
$p_y < 0$.

In the presence of the magnetic field ${\bf H}$ functions $u, v$ become spin-dependent,
and equations (\ref{HH})  are changed: $\nabla \to \nabla - i \frac{e}{c} {\bf A}$, $\mu \to
\mu_{\sigma} =\mu + \sigma \mu_B H$. For the constant magnetic field perpendicular to the plane
we have ${\bf A} = {\bf H} \times{ \bf r} /2$, so that $\partial_x \to \partial_x +i(e/c)yH/2$,
   $\partial_y \to \partial_y -i(e/c)xH/2$.
For the case $d_{x^2 -y^2}$ symmetry we consider $\Delta_{-\p}=
\Delta_{\p}=-\Delta_{\p -\Q}= 2(\cos p_x - \cos p_y ) \Delta_d
(\rr)$.

The self-consistent conditions
are derived  by substitution  of functions $u$, $v$ into (\ref{selfc}).
 In the continuum approximation
 they read:
\beq
\rho(\rr) = 2\sum_{\p} [(u_+^* u_+ + u_-^* u_- )f +
 (v_+^* v_+ + v_-^* v_- )(1-f)]
\label{s1}
\eeq
\beq (-1)^{x_i + y_i} m (\rr) = 4 U[\sum_{\p} u_-^* u_+ f - \sum_{\p} v_-^* v_+
(1-f)] \label{s2} \eeq
\beqa
\Delta_{\bf q}(\rr) =2g\sum_{\p} (v_{+}^* u_{+} -v_{-}^* u_{-})
[(1-f)(\cos (p_x-q_x)  \nonumber \\
+\cos (p_y-q_y) )-
f ((\cos (p_x + q_x) +\cos (p_y + q_y ) ) ],
\label{s3}
\eeqa
where $f = {1}/({\exp[\eps_{\p} /T] +1})$.
 We omitted spin indices since in our
representation for wave functions all equations are diagonal over spin.

\section{Superconductivity and spin-charge modulation}

In the absence of superconductivity (low doping limit)  the ground state  is
the periodic charge-spin superstructure  with $\Delta_d = 0$.  Spins are ordered antiferromagnetically
 ($S_z \propto (-1)^{x + y}$) for the nondoped  system ($\rho = 1$).  As a result of doping
 periodic structures of domain walls (stripe structure) are formed.
 Such solutions for the considered model were described  earlier \cite{fe,fe2}.
Below we consider region of doping that allows a superconducting phase .
In the general case a solution of the system of equations (\ref{HH}) is unknown.
We present  a solution taking into account both spin-charge and
superconducting structure  for a case of quasi-one-dimensional geometry, i.e. with broken fourfold rotational symmetry of the 2D-system.

There is a simplification
 for a  pure superconducting state ($m(\rr) = 0$), where
 equations (\ref{HH}) are reduced to:
\beq
-i \V_{\p} \nabla_{\rr} u_{\p} (\rr) + \Delta_{\p}v_{\p} = \eps u_{\p}
\label{v1}
\eeq
\beq
\Delta_{\p}^* u_{\p} + i\V_{\p} \nabla_{\rr}v_{\p} (\rr) =\eps v_{\p}.
\label{v2}
\eeq
The term $U\rho(\rr)/2$ in equations (\ref{HH}) can be eliminated by the shift
of wave functions $u,v \to u,v \exp i\Phi$, ${\bf V_p} \nabla \Phi
=U \rho(\rr )/2$ \cite{sm,mm}.
For filling factor $\rho  \sim 1$, the Fermi surface
has nearly square form, therefore
$\V_{\p} \nabla_{\rr} \approx V_{\p} \partial/\partial r_{\pm}$, depending on signs $p_x,  \, p_y$.
The system of equations (\ref{v1}), (\ref{v2})
has the following vortex solution:
\beq
\Delta_{\p}(\rr ) = \Delta_p\frac{\sinh \frac{r_+}{\xi_s} + i \sinh \frac{r_-}{\xi_s}}
{\sqrt{\sinh^2 \frac{r_+}{\xi_s} + \sinh^2 \frac{r_-}{\xi_s} + 1}},
\label{vortex}
\eeq
where $\Delta_p = \Delta_d (\cos(p_x) - \cos(p_y)) $; and $r_{\pm}=(x\pm y)/\sqrt{2}$.
For $r_- = 0$ the  order parameter has  a kink form $\Delta_p(\rr) \propto \tanh r_+ /\xi_s$,
where superconducting correlation length $\xi_s$ has to be determined by minimization of the kink's energy: $\xi_s\sim \bar{V_p}/\Delta_d$, and $\bar{V_p}$ is velocity averaged over the Fermi-surface.

For the case $r_+ = 0$ the order parameter acquires the phase:
$\Delta_p(\rr) \propto  \exp( i \pi / 2)\tanh r_- /\xi_s$.
In the  diagonal direction $r_+ = r_-$ the solution
$\Delta_p(\rr) \propto \exp(i \pi /4)\tanh (r_+ /\xi_s) /\sqrt{\tanh^2 r_+ /\xi_s  + 1} $ has the
  phase $\pi /4$.
It is known that  in one-dimensional case  finite-band solutions of equations (\ref{v1}) - (\ref{v2})
are related to the soliton (kink) solutions of the  nonlinear Schrodinger equation (NSE).
Note, that
along the curve $ \sinh r_- /\xi_s = \alpha \cosh r_+ /\xi_s$  the order parameter
acquires the form of a general kink solution of   the NES:
$\Delta_p(\rr) \sim (i \alpha + \tanh x/\xi_s )/\sqrt{\alpha^2 + 1}$ with  the localized state in the gap
with the energy $E_0 = \Delta_p \alpha /\sqrt{\alpha^2 + 1}$. We suppose  here that the magnetic field
 is small $H  \ll H_{c_2}$, ($\xi_s \ll \lambda$, $\lambda$ is magnetic penetration depth), and therefore ignore the effect of terms with vector-potential
  on the solution at distances $r \ll \lambda$.
The vortex structure (amplitude and phase) are shown in Fig. 2.
 \begin{figure}[tbph]
\begin{center}
\includegraphics[width=3.0in]{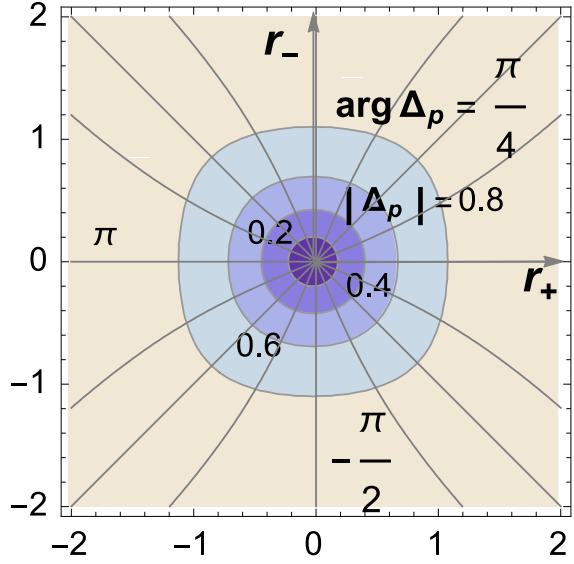}
\caption{(Color online) The superconducting vortex solution Eq. (\ref{vortex}) with contours of constant phase and modulus of the complex superconducting order parameter  $\Delta_{p}(\vec{r} )$.}
\end{center}
\label{vortex-a}
\end{figure}
For quasi-1D structures we can use  the ansatz \cite{ma}
\[
v_{\pm} = \gamma_{\pm} u_{\mp}.
\]
Considering $\eps(\p ) - \mu = 0$ on the Fermi surface
we obtain for the case
\[
m (\rr ) = \vert m(\rr) \vert e^{ i \varphi},\,
\Delta_{\p} (\rr) = \vert \Delta_{\p} (\rr)\vert e^{i\varphi_s},\,
\varphi,\, \varphi_s = const,
\]
the solution
$
\gamma_+ = \pm i e^{i(\varphi -\varphi_s )}, \,
\gamma_- = \pm i e^{-i(\varphi + \varphi_s)}$.
Equations  (\ref{HH})  are reduced to
\beqa
-i\V_{\p}\nabla u_+ + \tilde{\Delta}(\rr)u_- = E u_+\label{se0}\\
\tilde{\Delta}^*(\rr)u_+ + i\V_{\p}\nabla u_- = Eu_-
\label{se}
\eeqa
with function $\tilde{\Delta}(\rr) = (\vert m(\rr)\vert  \pm i\vert\Delta_{\p}\vert)
e^{i\varphi}$,  and    $m = m_{\pm}$, depending on the sign
of $p_y$. Equations (\ref{se0}), (\ref{se}) are exact provided that phases
$\varphi$,  $\varphi_s$ are constant or slowly varying in space functions.
We show that  inhomogeneity of the superconducting order parameter leads
to the origination of the antiferromagnetic order parameter.
Consider a 1D geometry  case: $ u = u(r_+)$, where assumption of constant phases is valid.
 The  solution of Eqs. (\ref{se0}), (\ref{se}) describing the coexistence of superconductivity
   and spin-charge  density ordering,  compatible  with self-consistent equations,
 has the form of two bound  solitons of the nonlinear Schrodinger equation, see, for example, \cite{fadtakh}
 \beq
 \tilde{\Delta}_{1,2} = \Delta_p \frac{\cosh (2 \kappa x + c_1) + \cosh (c_2 \pm 2 \beta i)/|\lambda| }
 {\cosh (2 \kappa x + c_1) + \cosh (c_2 )/|\lambda|},
 \label{d12}
 \eeq
 where $E =\pm \lambda ({\bf p})$ are positions of local levels inside the gap,
 $\kappa = \kappa_p =\sqrt{\Delta_p^2 - \lambda^2}$, and $\exp i\beta = \lambda + i\kappa$.
 Eigenfunctions of equations (\ref{se0}), (\ref{se})  have the form \cite{bm}
 \beq
 u_{\pm}(x) \propto \sqrt{\tilde{\Delta}(x) (E^2 - \gamma^2(x))}\exp \left[\pm i\int^x
 \frac{\sqrt{(E^2 - \Delta_{p}^2)(E^2 - \lambda^2)}}{E^2 - \gamma^2(y)} dy \right],
 \label{wavefunction}
 \eeq
 where
 \beq
 \gamma (x) = \frac{1}{2} \frac{\partial}{\partial x} \ln \tilde{\Delta}(x).
 \eeq
The real and imaginary parts of (\ref{d12}) give
superconducting and spin order parameters
\beq
\Delta_{sc} = \Delta_p (1 - \Gamma \tanh a (\tanh(\frac{r_+}{\xi} + \frac{a}{2}) - \tanh (\frac{r_+}{\xi}- \frac{a}{2} ))),
\label{scorder}
\eeq
\beq
m = m_0 \Gamma \tanh a (\tanh(\frac{r_+}{\xi} + \frac{a}{2}) - \tanh (\frac{r_+}{\xi}- \frac{a}{2} )),
\label{morder}
\eeq
where we use a parametrization: $ \Delta_p \sqrt{\Gamma_p} \tanh a = \kappa_p$,  and
 $ \Gamma_p=\Delta^2_p /(\Delta_p^2 + m_0^2)$.
 Averaged over Fermi surface functions are defined as:  $\Gamma = <\Gamma_p>=<\Delta^2_p /(\Delta_p^2 + m_0^2)>_p $,  $\xi = <v_p/(\Delta_p \sqrt{\Gamma_p} \tanh a)>_p$.
 Values of $m_0$,  $\Delta_{\p} = \Delta_d (\cos p_x - \cos p_y)$,
$\xi$ and a dimensionless parameter $a$ are defined by
the self-consistency conditions (\ref{s2}), (\ref{s3}); $m_0\propto t\times exp\{-t/U\}$
The solution describes the spin-charge stripe in superconducting phase  (see Fig.3).

 \begin{figure}[tbph]
\begin{center}
\includegraphics[width=3.0in]{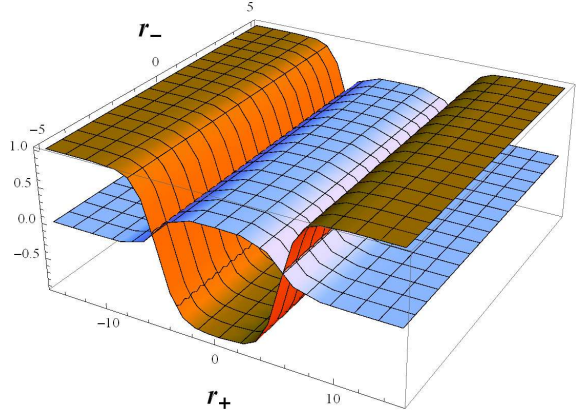}
\caption{(Color online) Coexisting superconducting (downward) and antiferromagnetic stripe-like (upward) orders. The envelope functions from Eqs. (\ref{scorder}), (\ref{morder}) are plotted in real space, $r_{-}$ and $r_{+}$ coordinates are measured  in units of correlation length $\xi$, other parameters are: $a=7$, $\Delta_d=1$, $m_0=0.2$.}
\end{center}
\end{figure}

The spin inhomogeneity generates the charge distribution $\delta \rho (\rr ) \propto
m^2(\rr )$. Note, that two-soliton solution in the similar form was used for describing
polaron-bipolaron states in the Peierls dielectrics \cite{brkir}.

The superconducting  correlation length  is increased in comparison with clean superconductor case $\xi_{s}$  as $\xi = \xi_{s} \sqrt{1+ \xi^2_{s} /\xi^2_{AF}}$, where the following definitions are assumed: $\xi_{s}=<v_p/\Delta_p >_p$, and $\xi_{AF}=<v_p/m_0 >_p$.

The obtained structure is stabilized by applying a magnetic field. To show this we  include the
uniform magnetic field, perpendicular to the plane of the system.
We will work  in the limit of extreme
type-II superconductivity (with Ginzburg-Landau
parameter   $\kappa = \lambda/\xi  \to  \infty $, $H_{c1} \sim \ln \kappa /\kappa \to 0$); so there is no screening of the magnetic
field by the Meissner currents, and $\nabla  \times {\bf A} = H  \hat{ \bf z}$,
is an applied, space-independent magnetic field.
In the symmetric gauge we have $ A_x = - y H/2$, $A_y = x H/2$.  In equations (\ref{se0}), (\ref{se}) we
substitute $-i\nabla \to -i\nabla - e{\bf A}/c$.
The magnetic field can be excluded by transformation $u, v \to ( u, v) \exp (i e H r_+ r_- /2c)$.

The energy of the stripe structure (\ref{scorder}), (\ref{morder}) in the magnetic field is given by
\beq
W = \sum E_j + \int d^2\\r \left[\frac{\Delta_d^2}{g^2}+ \frac{m^2}{g_{AF}^2} \right] -\int M H d^2 r -W(a=0, m=0),
\label{ws}
\eeq
where
\beqa
-\int M H d^2 r =  - \int m(r_+) \cos(Q  r_+) H d r_+ = \nonumber\\
-\frac{ m_0 \Gamma H   \xi  \tanh a}{\pi v_p \lambda}\frac{2 \pi \sin (Q \xi a /2)}{\sinh (\pi Q \xi /2)}=
  - m_0 H \frac{2}{\lambda \Delta_p} \sqrt{\Gamma} \frac{\sin\frac{Q v_F a}{2\sqrt{\Gamma} \Delta_p \tanh a}}
{\sinh \frac{\pi Q v_F}{2 \sqrt{\Gamma} \Delta_p \tanh a}},
\label{mH}
\eeqa
$\xi  = \xi_0 /\tanh a$, $\xi_0 = v_F / (\Delta_p \sqrt{\Gamma})$.


The first sum  in (\ref{ws}) is found similar  to \cite{brkir}.
After  cumbersome calculations   we obtain

\beqa
W/L = \frac{4}{\pi}\Delta_p \left[    \sqrt{\Gamma} \tanh a - \sqrt{1 - \Gamma \tanh^2 a}\,\arcsin [  \sqrt{\Gamma} \tanh a]\right]
+ \frac{4}{\pi} m_0^2 \frac{\sqrt{\Gamma}\tanh a}{\Delta_p \, \lambda} + \nonumber \\
 \frac{4}{\pi} m_0^2 \frac{\Gamma^{3/2}}{\Delta_p } (a - \tanh a)\left(\frac{1}{\lambda_{AF}} - \frac{1}{\lambda} \right) - m_0 H \frac{2}{\Delta_p \lambda} \sqrt{\Gamma} \frac{\sin\frac{Q v_F a}{2\sqrt{\Gamma} \Delta_p \tanh a}}
{\sinh \frac{\pi Q v_F}{2 \sqrt{\Gamma} \Delta_p \tanh a}},
\eeqa
 where $\lambda = g^2/\pi v_F$.
A typical dependence $W(a)$ is shown in Fig. 4

 \begin{figure}[tbph]
\begin{center}
\includegraphics[width=3.0in]{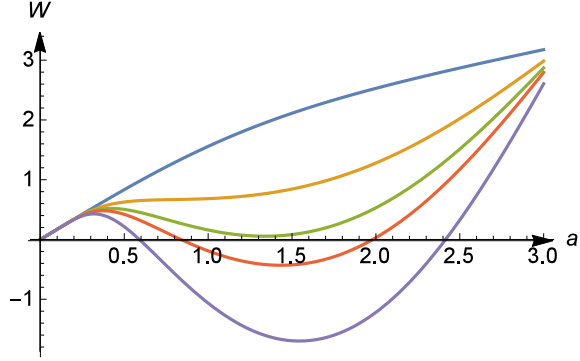}
\caption{(Color online) The energy W(a)  for magnetic fields $h =0$ (top curve), $h = 0.5, 0.8, 1$ and $h = 2$ (bottom curve)  }
\end{center}
\end{figure}
 The  energy has a minimum with $W(a_0) < 0$   at $a_0 \sim 1$  in a magnetic fields $\mu_B H \gtrsim m_0$ .

We conclude that in the region of inhomogeneity of superconducting order parameter a spin-density is formed.
This coexisting structure is stabilized by applied magnetic field due to the fact, that $-\int M H d^2 r$ in Eq.(\ref{mH}) over a finite-size AFM stripe is nonzero due to envelope shape suppression AFM spin variation close to the boundaries of the AFM stripe.

Note, that a similar solution  can be written for the  low-doping AFM spin-density wave region.  The applied magnetic
 field stabilizes the stripe structure and results in generation of superconducting order parameter. Hence, the system
  possesses a kind of 'SO(5)' symmetry.

\section{Discussion}

We considered a simple self-consistent 2D model  on a squared lattice to describe different
states, including charge-spin structures, superconductivity, and their coexistence.
The origin of spin-charge periodic state (which is responsible for the pseudogap) is due to the existence of flat parallel
segments of the Fermi surface (nesting) at low hole doping concentrations.
Effects of commensurability
lead to a pinning of stripe structure at rational filling
points $|\rho - 1| = m/n$.
As a result,
there is  an exponentially small (for large $n$) decrease in the total energy of the order
$\delta E \sim \exp( - c\, n)$   at any commensurate point, stabilizing stripes, as in 1D systems.
For this reason, we think, stripes  are mostly observable  near $n = 8$ point ($|\rho- 1|= 1/8$).
An increase of doping leads to the decrease of flat segments of the Fermi surface and attenuation of spin-charge structure.

We found the solution describing the coexistence of superconductivity and stripes (\ref{scorder}), (\ref{morder}).
The decrease (or a deviation from the homogenous value) of the superconducting order parameter
generates the spin-charge periodic structure in this region.  Note, that due to symmetry
of Eqs. (\ref{HH}),(\ref{se0}), (\ref{se})  (duality $\Delta \leftrightarrow i m$)
we can write the same equation, describing the origin of superconducting  correlations in the region
of a inhomogeneity of spin-charge density.  The situation is qualitatively similar to the 1D case \cite{ma}.
Experimental data in underdoped high-T$_c$ cuprates LSCO \cite{aeppli1} indicates that antiferromagnetic stripe-like
spin-density order can be induced by magnetic field perpendicular to the CuO planes in the interval of fields much smaller than
upper critical field H$_{c2}$ . The size of the magnetically ordered domains exceeds superconducting vortex's core size $\xi_s$ and the
inter-vortex distance in the Abrikosov's lattice.    Our present theoretical results demonstrate that this is indeed possible in the simple Hubbard
t-U-V model that we consider. In particular, the dimensionless parameter $a$ in Eqs. (\ref{scorder}), (\ref{morder})  is an independent variational parameter and depends
on the magnetic and superconducting coupling strengths \cite{ma}, as well as on the magnitude of the external magnetic field. Hence, the size $\sim a\times \xi$ of the
antiferromagnetic domain (see Fig. 3, upward bump) can exceed the superconducting (and magnetic) Ginzburg-Landau correlation length $\xi$
when $a(H)>>1$. Previously coexistence of superconducting order and slow antiferromagnetic fluctuations was studied merely on the basis of a phenomenological Ginzburg-Landau free energy functional approach in \cite{sachdev1}.
We note also, that equations Eqs. (\ref{se0}), (\ref{se}) can be simply extended to include d-density waves (DDW).
\\
When this work was already done, we found a recent paper, in which authors \cite{efetov} had provided theoretical
description of the onset of charge ordered domain in the vortex core, where the superconducting order parameter
 turns to zero. This approach differs from ours in three major respects. The first one is that in \cite{efetov} the absence
  of antiferromagnetic (AFM) order is assumed, while critical antiferromagnetic fluctuations are considered to provide
  coupling between electron-hole and Cooper channel fluctuations at the hot-spots and antipodal regions of the
  under-doped cuprates Fermi-surface, leading to a coexistence of superconducting (SC) and charge orders (CDW).
 The influence of an external magnetic field is considered in \cite{efetov} merely as the prerequisite of CDW formation
 inside the superconducting cores. The second important respect of the difference with our approach is that the
  authors \cite{efetov} consider an SU(2)-symmetric  composite order parameter (CDW and SC), while we consider
   analytically a composite CDW-AFM-SC order parameter, since the AFM 'stripe' coexisting with superconductivity
  incorporates CDW, AFM and SC orders. Finally, the third aspect of the difference is that the external magnetic field
  in our model couples to  both the SC order and to finite-width stripe of AFM order. Unlike in the CDW-SC coexistence
   case in \cite{efetov}, this leads to a two-fold stabilization of the stripe phase inside the superconducting cores:
  both by suppression of superconductivity and by dipole-like coupling of a net magnetic moment of a finite size AFM
   domains to magnetic field. Definitely, our model yet lacks all the versatile families of hot spot and antinodal nesting vectors
  (to be farther allowed for in the future), as considered in \cite{efetov}, while so far we considered only two
   anti-ferromagnetic nesting wave-vectors instead.
\\

\section{Acknowledgements}
 The work was carried out with financial support in part from the Ministry of Education and Science of the Russian
  Federation  in the framework of Increase Competitiveness Program of NUST "MISIS"  (No. K2-2014-015)  and from RFFI grant 12-02-01018.


\end{document}